# Epitaxially Strained BiMnO$_3$ Films: High-Temperature Robust Multiferroic Materials with Novel Magnetoelectric Coupling

X. Z. Lu[a)], X. G. Gong, and H. J. Xiang[b)]

Key Laboratory of Computational Physical Sciences (Ministry of Education), State Key Laboratory of Surface Physics, and Department of Physics, Fudan University, Shanghai 200433, P. R. China

**ABSTRACT**

Multiferroics with the coexistence of ferroelectric and ferromagnetic orders are ideal candidates for magnetoelectric applications. Unfortunately, only very few ferroelectric-ferromagnetic multiferroics (with low magnetic critical temperature) were discovered. Here we perform first principles calculations to investigate the effects of the epitaxial strain on the properties of BiMnO$_3$ films grown along the pseudocubic [001] direction. Unlike the ground state with the centrosymmetric $C2/c$ space group in bulk, we reveal that the tensile epitaxial strain stabilizes the ferromagnetic and ferroelectric $Cc$ state with a large polarization ($P > 80$ μC/cm$^2$) and high Curie temperature ($T_c$ is predicted to be between 169 K and 395 K). More importantly, there is a novel intrinsic magnetoelectric coupling in the multiferroic $Cc$ state with the easy magnetization axis controllable by the external electric field.

a) 11210190025@fudan.edu.cn; b) hxiang@fudan.edu.cn

## I. INTRODUCTION

In recent years, multiferroics, displaying magnetic, polar, and elastic order parameters simultaneously, have attracted numerous research interests[1]. Among various kinds of multiferroics, multiferroic materials with coexisting ferromagnetism and ferroelectricity, i.e., ferroelectromagnets, are particularly intriguing because they offer the best prospect of achieving a strong linear ME coupling coefficient owing to the combination of typically higher electric permittivity and magnetic permeability and thus may lead to the development of multistate logic, new memory or advanced sensor technologies. Unfortunately, only very few ferromagnetic (FM) and ferroelectric (FE) multiferroics were discovered: $Eu_{1/2}Ba_{1/2}TiO_3$ is the first compound exhibiting simultaneously ferroelectricity and ferromagnetism below 4.2 K[2,3]; $EuTiO_3$ under large biaxial strain could be a FE ferromagnet with a low FM Curie temperature 4.24 K[4]; It was predicted that $SrMnO_3$ could be driven by epitaxial strain to a multiferroic FM-FE state[5] with the estimated magnetic Curie temperature around 92 K within the mean field theory. $BiFeO_3$ is a well-known room temperature multiferroic material, but it is antiferromagnetic with a modulated cycloidal spin structure[6]. Another recently reported room temperature multiferroic compound is $Sr_3Co_2Fe_{24}O_{41}$ with the FE ordering induced by helimagnetic ordering[7]. For realistic applications, it is desirable to discover or design a room-temperature FM-FE multiferroics with a strong coupling between FE and FM orders.

Among various candidates for FM-FE multiferroics, perovskite-type $BiMnO_3$ is special because of the initially predicted coexistence of ferromagnetic and ferroelectric order in this system[8]. Although there was no doubt on its ferromagnetism[9-16], controversies on its ferroelectricity arose from the most recent experimental and theoretical investigations[15-18], which demonstrated that $BiMnO_3$ crystallizes in a centrosymmetric space group $C2/c$. Therefore, it is most likely that there is no ferroelectricity in bulk $BiMnO_3$. In this work, we propose to tune the magnetic and polar order in $BiMnO_3$ thin films by epitaxial strain, which is motivated by the recent studies on epitaxial perovskite thin films[4,5,19-23] showing that strain could tune the magnetic and polar order. To the best of our knowledge, the properties of the epitaxial pseudocubic [001]-oriented $BiMnO_3$ thin film are not fully understood both experimentally and theoretically.

In this paper, we carry out comprehensive first principles studies to investigate the effects of the epitaxial strain on the properties of $BiMnO_3$ films grown on the cubic substrate along the pseudocubic [001] direction. A series of phase transitions occur in the studied strain region such as a transition between the AFM insulating FE phase and FM metallic phase and a transition between the FM-PE phase and FM-FE phase. Furthermore, there is a novel intrinsic magnetoelectric coupling in the FM-FE $Cc$ state with the calculated polarization $P > 80$ µC/cm$^2$ and estimated Curie temperature $T_c$ is between 169 K and 395 K.

## II. COMPUTATIONAL METHODS

Total energy calculations are based on the density functional theory (DFT) plus the on-site repulsion (U) method[24] within the generalized gradient approximation with the revised Perdew-Becke-Erzenhof (PBEsol) parametrization[25] on the basis of the projector augmented wave method[26] encoded in the Vienna ab initio simulation package[27]. The plane-wave cutoff energy is set to 700 eV and $4 \times 4 \times 3$ $k$-point mesh is used for the 20-atom $\sqrt{2} \times \sqrt{2} \times 2$ cell. We discuss the results obtained with the on-site repulsion U = 4 eV (see Part 1 of Ref. 28) and the exchange parameter J = 1 eV on Mn. For the calculation of electric polarization, the Berry phase method[29] is used unless noted otherwise.

The lattice vectors used in this study are $\mathbf{a} = a_s\mathbf{x} - a_s\mathbf{y}$, $\mathbf{b} = a_s\mathbf{x} + a_s\mathbf{y}$ and $\mathbf{c} = \delta_1\mathbf{x} + \delta_2\mathbf{y} + (2a_s + \delta_3)\mathbf{z}$, where $a_s$ is the pseudocubic lattice constant of the studied system at a given strain, and ($\mathbf{x}$, $\mathbf{y}$, $\mathbf{z}$) are defined within the pseudocubic setting. Epitaxial strain is then defined as $(a_s - a_0)/a_0$, where $a_0$ (3.938 Å in this study) is assumed to be the pseudocubic lattice constant of bulk structure with *C*2/*c* space group under a volume-conserving transformation. According to the previously reported possible stable perovskite structures under epitaxial strain[5,20-23,30-32], we have investigated the BiMnO$_3$ thin films under coherent epitaxial strain (i.e, both the length and direction the in-plane lattice vectors are fixed to those of a cubic substrate) in the *I*4/*mcm*, *Imma*, *R*$\bar{3}$*c*, *C*2/*m*, *Pnma*, *Ima*2, *C*2, *Cc*, *Pmc*2$_1$ space groups with the FM, A-type AFM (A-AFM), G-type AFM (G-AFM), C-type AFM (C-AFM) spin orders. To obtain the initial structures of the different space groups, we use the method described by Stokes *et al.*[33]. In order to obtain the minimum-energy configuration in a given space group, we optimize the length and direction of the out-of-plane lattice vector (*c* lattice vector) and the internal ionic positions until Hellman-Feynman forces are less than 0.01 eV/Å. For the property calculations, we focus on the lowest energy phases in the studied strain range. Curie ($T_c$) and Néel ($T_N$) temperatures are estimated using mean field theory and Monte Carlo simulations (see Part 2 of Ref. 28), in which the in-plane nearest-neighbor ($J_1$) and out-of-plane nearest-neighbor ($J_2$) spin exchange interactions are considered and extracted from the energy differences between the FM, A-AFM, G-AFM, C-AFM orders (see Part 3 of Ref. 28).

## III. RESULTS AND DISCUSSION

The total energies of the lowest energy phases, as a function of epitaxial strain from −6.2% to 7.8%, are shown in Figure 1. For any strain, the G-AFM and C-AFM states are in much higher energy than the FM and A-AFM states. Thus, they are not shown in Figure 1. For comparison, we also apply the coherent epitaxial strain to the $C2/c$ structure with 80 atoms with the strain ranging from -3.2% to 3.8%. For this structure, the lattice vectors are $\mathbf{a} = 2(a_s\mathbf{x} - a_s\mathbf{y})$, $\mathbf{b} = a_s\mathbf{x} + a_s\mathbf{y}$ and $\mathbf{c} = \delta_1\mathbf{x} + \delta_2\mathbf{y} + (4a_s + \delta_3)\mathbf{z}$[34]. The results are shown in Figure 1. The electric polarizations, magnetic Curie/Néel temperatures, band gaps and $c$-lattice parameters for these lowest energy phases are shown in Figure 2. Our results, as discussed in the followings, are different from those obtained by Hatt et al.[35], who examined the effects of strain on BiMnO$_3$ by applying the biaxial strain with a fixed ratio ($a/b$ = 1.01) between the two pseudocubic in-plane lattice constants (not coherent epitaxial strain) to the experimental $C2/c$ structure and concluded that BiMnO$_3$ remains centrosymmetric and non-polar in the strain range of -3 – 3% (being equivalent to -4.5

– 1.4% in our study).

As can be seen in Figure 1, if the strain is small, i.e., between -0.7% and 2.8%, the orthorhombic centrosymmetric *Pnma* structure has the lowest energy, for which the FM order is the spin ground state. Although the space group of BiMnO$_3$ has been changed from *C*2/*c* to *Pnma*, the unusual FM order remains to be the spin ground state in contrast to other perovskite materials where usually the AFM states prevail, e.g., LaMnO$_3$ with the same space group has the A-AFM spin order. The FM order is further confirmed by the calculated spin exchange parameters (see Part 3 of Ref. 28). Interestingly, the *Pnma* structure was also found in the polycrystalline BiMnO$_3$ samples at high temperature[12]. With increasing strain, the FM *Pnma* structure becomes unstable, with a first-order phase transition into the FM *Cc* phase.

When the compressive strain is larger than 0.7%, the FM *Pnma* structure becomes unstable and the FM *Cc* structure is favored. This FM *Cc* phase is metallic, thus not FE. It should be noted that both *Pnma* and *Cc* phases were found in the simulations of epitaxial strained [-110]-oriented BiFeO$_3$ thin film[22]. Interestingly, a phase transition from the FM metallic (FM-*M*) state to the A-AFM insulating FE (A-AFM-*I*-FE) state occurs in the *Cc* structure at -5.2% strain[36]. A similar phenomenon was also observed in a recent investigation on the epitaxial strained SrCoO$_3$[21]. This A-AFM-*Cc* state with a band gap larger than 0.2 eV displays a large FE polarization and a relatively high $T_N$ of ~304 K. As discussed in Ref. 21, in the vicinity of this phase transition, the magnetic-electric-elastic response may be expected due to the changes of the physical properties such as FE polarization, magnetic order, band gap and out-of-plane lattice parameter between the two phases. Thus we predict a possibility of a metal-insulator transition at a compressive strain in BiMnO$_3$ thin films grown along the pseudocubic [001] direction. For examples, the A-AFM-*I*-FE phase can be transformed into the FM-*M* phase by applying magnetic field or a compressive uniaxial stress that will lead to an insulator-metal transition, while the opposite transition can be induced by applying electric field or a tensile uniaxial stress.

In the tensile strain region, there is a phase transition at 2.8% strain from the PE-FM *Pnma* state to FE-FM *Cc* state. Although it is hard to tell whether the A-AFM *Cc* state has lower energy than the FM *Cc* state in Figure 1, the total energy and extracted out-of-plane spin exchange parameters (see Part 3 of Ref. 28) provide unambiguous evidence that the FM order is preferred. In particular, the FM order is much more stable than the A-AFM order within the 2.8–3.8% strain region. The magnetic Curie temperature for the FM *Cc* state at 2.8% strain is estimated to be $T_c$~395K within the mean field approximation. We find that the FM *Cc* state with the band gap of ~ 0.12 eV has an electric polarization as large as 80 μC/cm$^2$. As expected from the polarization-strain coupling characteristic of the ferroelectric perovskites, the in-plane polarization ($P_{x/y}$) of FM-*Cc* state increases with the increasing strain whereas the out-of-plane polarization ($P_z$) decreases, while the opposite behaviors are found for A-AFM *Cc* state in the compressive region (see Figure 2a). Therefore, we predict that the tensile strain could stabilize a high temperature FM-FE phase in BiMnO$_3$ thin film. Interestingly, the predicted low temperature FM-FE phase in the tensile strain range in EuTiO$_3$ thin films was confirmed by a recent experiment[4]. Although the transition occurs at a large strain at 0 K in our study, the strain may be reduced at a finite temperature according to the reports on EuTiO$_3$ thin films[4].

For all the new three phases (A-AFM-FE *Cc*, FM-PE *Pnma* and FM-FE *Cc* phases) of BiMnO$_3$ predicted in this work, we find that each Mn ion has two long Mn-O bonds along the local z axis and four short Mn-O bonds in the local xy-plane. Thus, the Mn$^{3+}$ ion (d$^4$) has the $t_{2g}^3 d_{z^2}^1$ electron configuration. Indeed, the partial charge distribution (Figures 3a-3c) shows that the highest occupied *d* state is $d_{z^2}$-like. Interestingly, we find that the orbital orders of Mn ions for the three phases are different from each other. In the A-AFM-FE *Cc* phase, the Mn ions have an occupied $d_z^2$ orbital along the *c*-axis, while the occupied $d_z^2$ orbitals of Mn ions lie in the *ab*-plane for the FM-PE *Pnma* phase where the $d_z^2$ orbital is approximately orthogonal to the in-plane nearest neighbors, but parallel to the out-of-plane nearest neighbors, similar to the LaMnO$_3$ case. As for the FM-FE *Cc* phase, it has a similar in-plane arrangement of the $d_z^2$ orbital as the FM-PE *Pnma* phase, but an almost orthogonal arrangement between the adjacent planes along the *c*-axis. The orbital orders lead to the magnetic ground states of the three phases, as shown in the fourth part of Ref. 28.

In order to study why the *Cc* structure can generate such a large FE polarization, we also calculate the polarization by summing the product of the ions displacements with the Born effective charges (see Part 5 of Ref. 28) in the *Cc* structure at 3.8% strain, which is consistent with the result from the direct Berry phase calculation. The reference paraelectric structure (*C*2/*c* space group) is obtained by adding the centrosymmetric operation to the *Cc* structure. Then we find that the contributions of $Bi^{3+}$, $Mn^{3+}$, $O^{2-}$ ions to the total polarization are (30.762, 28.043, 27.291), (0.684, -0.957, -0.061), (18.887, 23.227, 10.168) $\mu C/cm^2$, respectively. So, this large polarization is mainly due to the displacements of $Bi^{3+}$, $O^{2-}$ ions. To find out the origins of these displacements, we calculate the electron localization function for the state. As shown in Figure 3d, the $6s^2$ lone pairs are distributed along the same directions, which is different from the situation of *C2/c* bulk structure in which the $Mn^{3+}$ ions arranged along the *b*-axis have the opposite distributions of $6s^2$ lone pairs viewed from the *c*-axis[17]. Therefore the $Bi^{3+}$ ions move along the same directions. The final displacements of $Bi^{3+}$, $O^{2-}$ ions are found to be along the pseudocubic [11z] direction because of the existence of glide plane lying in the (1-10)-plane (*bc*-plane in Figure 3d), which results in the [11z] direction for the FE polarization. The ferroelectricity induced by the epitaxial strain in the *Cc* structure may be due to the long-range Coulomb interactions related to the Bi ions, different from the strained $CaMnO_3$ case where the $Mn^{4+}$ ions are responsible for the ferroelectricity[19].

So far, we have predicted the coexistence of the ferromagnetism and ferroelectricity in the *Cc* phase (strain > 2.8%). The FM-FE *Cc* structure is also found to be dynamically stable (see Figure S6 of Ref. 28). And the ground state nature of the FM-FE *Cc* structure is further confirmed by our global optimization genetic algorithm simulations[37]. Another important question is that whether there exists magnetoelectric coupling between the two ferroic orders in the phase. First, we calculate the single-ion anisotropy (SIA) of $Mn^{3+}$ ion in the *Cc* structure at 3.8% strain using DFT+U+SOC method. The results show that the $Mn^{3+}$ ion has an easy-axis anisotropy with the local easy-axis close to the longest Mn-O bond direction, as shown in Figure 3d. The direction of the easy axis can be explained by the second order perturbation theory, similar to the $TbMnO_3$ case[38]. As a result, the easy magnetization axis of the FM state will be nearly in the *ab*-plane because all four local easy axes are almost in the plane. The total SIA energy of the FM *Cc* state can be written as $E_{SIA} = \sum_i A(\mathbf{S}_i \cdot \mathbf{n}_i)^2$, where $A = -3.105$ meV/Mn is the effective single-ion anisotropic parameter, $\mathbf{S}_i = \mathbf{S}(\theta, \varphi)$ represents the i-th $Mn^{3+}$ spin direction ($\theta$ and $\varphi$ are the zenith and azimuth angles of the magnetization direction defined in the pseudocubic coordination system), and $\mathbf{n}_i$ is the direction of the local easy-axis. We then find that the total SIA energy is the lowest when $\theta = 90°$ and $\varphi = 135°$, i.e., the pseudocubic [1-10] direction (see Part 7 of Ref. 28). We can see that the easy magnetization axis is orthogonal to the direction of the electric polarization. The hard magnetization axis is along the direction with $\theta \approx 165°$ and $\varphi = 45°$. The energy difference between the hard magnetization case and the easy magnetization case is as large as $|A|/2 \approx 1.55$ meV/Mn. When the spins are in the *ab*-plane ($\theta = 90°$), $E_{SIA}(\phi) \sim B\sin(2\phi)$ ($B > 0$). The $\sin(2\phi)$ dependence of the total energy is in good agreement with the results from the direct DFT+U+SOC calculations (Figure 4a square dots).

Because the low symmetry of the *Cc* structure, we would expect that there will exist eight different FE domains with different polarizations $P = (\pm x_0, \pm x_0, \pm z_0)$[39]. The easy magnetization axis of the four FE domains $[P = (x_0, x_0, z_0), P = (x_0, x_0, -z_0), P = (-x_0, -x_0, z_0), P = (-x_0, -x_0, -z_0)]$ will be almost orthogonal to that of the other four domains $[P = (x_0, -x_0, z_0), P = (x_0, -x_0, -z_0), P = (-x_0, x_0, z_0), P = (-x_0, x_0, -z_0)]$. From the above discussions, we propose that there should exist intrinsic magnetoelectric coupling in the FM-FE *Cc* phase: The two types of domains can be switched by the external electric field, thus the $Mn^{3+}$ spins directions can be controlled by an electric field (see Figure 4b). The control of the magnetic order by the electric field was predicted in previous first principles calculations[4,5,21], in which it occurs at the phase boundary between two phases with different magnetic and polar orders. Experimentally[40,41], a ferroelectric control of spin direction was also demonstrated: The control of the spin direction arises from the pure interface effect where the switch of the FE polarization of the bottom ferroelectric layer induces the changes of the magnetic anisotropy of the top layer. However, the strong magnetoelectric coupling due to the strong SIA of Mn ion predicted in this FM-FE *Cc* phase is intrinsic. The mechanism for manipulating magnetization vector by electric fields proposed in this work is also different from other cases: in the diluted magnetic semiconductor case [e.g., (Ga,Mn)As], where an electric field tunes the hole concentration and thus the magnetic anisotropy[42]; in epitaxial (001) $BiFeO_3$ films, where the easy magnetization plane (antiferromagnetic) can be switched by the 71° or 109° ferroelectric domains switching at room temperature[43]; in $(Cu;Ni)B_2O_4$ compound, it was the modulation (up to ±30°) of the weak magnetization (≈ 0.006 $\mu_B$/f.u. due to the Dzyaloshinskii-Moriya interactions) by the external electric field in a canted-antiferromagnetic phase at 15K[44]. Furthermore, the novel intrinsic magnetoelectric coupling revealed in this study does not require the same magnetic ion to be responsible for both ferroelectricity and magnetism, suggesting a new direction for searching new multiferroics. We note that the magnetoelectric coupling due to the SIA may be important in systems with anisotropic d orbital occupation (e.g., $d^4$ in $BiMnO_3$) but not in isotropic Heisenberg systems (such as $EuTiO_3$ and $SrMnO_3$). Besides, a very recent experiment[45] by using a planar electrode device can directly measure the in-plane component of polarization in a fully strained $BiFeO_3$ thin film at a compressive strain as large as 4.4%, which may also be employed to study other similar systems such as $BiMnO_3$. So we hope that our theoretical work can stimulate future experimental works on strained $BiMnO_3$.

As is well known, the mean field theory usually overestimates the critical temperature of the magnetic order. Therefore, we also estimate the critical temperature by Monte Carlo simulations (see Part 2 of Ref. 28). The results show that the $T_c$ at 2.8% strain and $T_N$ at -5.2% strain are 169 K and 142 K, respectively. Since the classic Monte Carlo simulation usually underestimates the critical temperature, we expect that the critical temperatures for $T_c$ and $T_N$ may be between the two values obtained from the mean field theory and Monte Carlo simulations, respectively.

## IV. CONCLUSIONS

In summary, we performed first principles calculations to investigate the effects of the epitaxial strain on the properties of $BiMnO_3$ films grown along the pseudocubic [001] direction. Unlike the ground state of the centrosymmetric $C2/c$ space group in bulk, three previously unreported phases, namely, A-AFM *Cc*, FM *Pnma* and FM *Cc* phases, are stabilized under epitaxial strain. For tensile strain larger than 2.8%, the *Cc* structure has a large electric polarization ($P > 80$ μC/cm$^2$) and is FM with a large saturated magnetic moment (4 $\mu_B$/Mn) and a high magnetic Curie temperature (estimated $T_c$ is predicted to be between 169 K and 395 K). To the best of our knowledge, BiMnO$_3$ under tensile strain has the highest Curie temperature among all FM-FE multiferroics. Moreover, a novel intrinsic magetoelectric coupling is predicted to exist in the FM-FE *Cc* phase with the easy axis tunable by the external electric field, which indicates the potential application of epitaxial (001) BiMnO$_3$ thin films in spintronic devices.

**ACKNOWLEDGEMENTS**

Work at Fudan was partially supported by NSFC, the Special Funds for Major State Basic Research, Foundation for the Author of National Excellent Doctoral Dissertation of China, The Program for Professor of Special Appointment at Shanghai Institutions of Higher Learning, Research Program of Shanghai municipality and MOE.

(34) When the epitaxial strain is imposed on the bulk FM *C*2/*c* structure, the symmetry is changed to *P*$\bar{1}$. After optimizing the FM *P*$\bar{1}$ structures under the epitaxial constraint, the FM *P*$\bar{1}$ structure automatically changes to the FM *C*2/*c* structure in the range of -3.2 – 0.8%, while it keeps the *P*$\bar{1}$ symmetry in the range of 1.8 – 3.8%. Therefore, we refer this special phase as "epitaxial FM $C2/c$" phase.

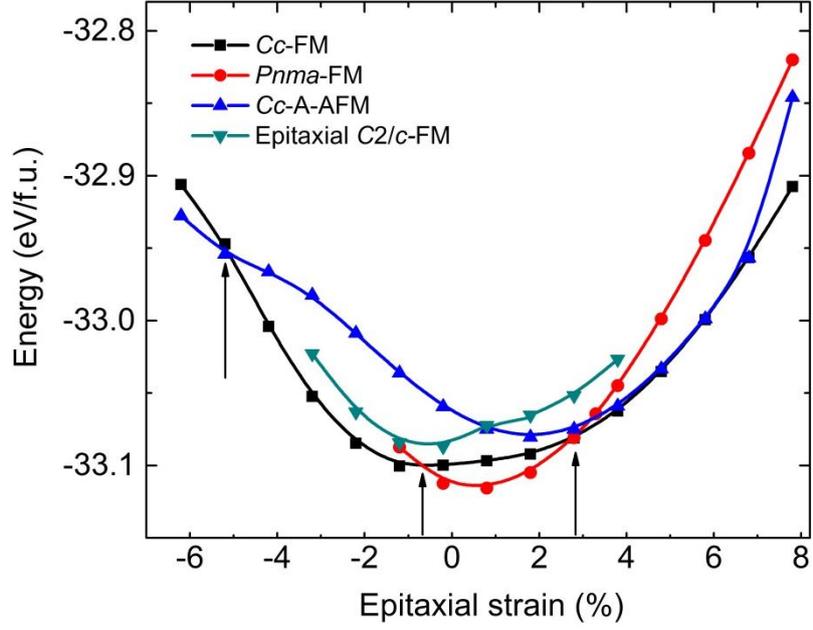

FIG. 1. Calculated total energies versus the epitaxial strains for the various spin orders with *Pnma* and *Cc* space groups. The total energies of the epitaxial FM *C*2/*c* phase[33] are also plotted over the strain range from -3.2% to 3.8% for comparison. The vertical black lines at -5.2%, -0.7% and 2.8% indicate the phase boundaries between A-AFM-*I*-FE *Cc*, FM-*M Cc*, FM-*I*-PE *Pnma*, and FM-*I*-FE *Cc* phases, respectively.

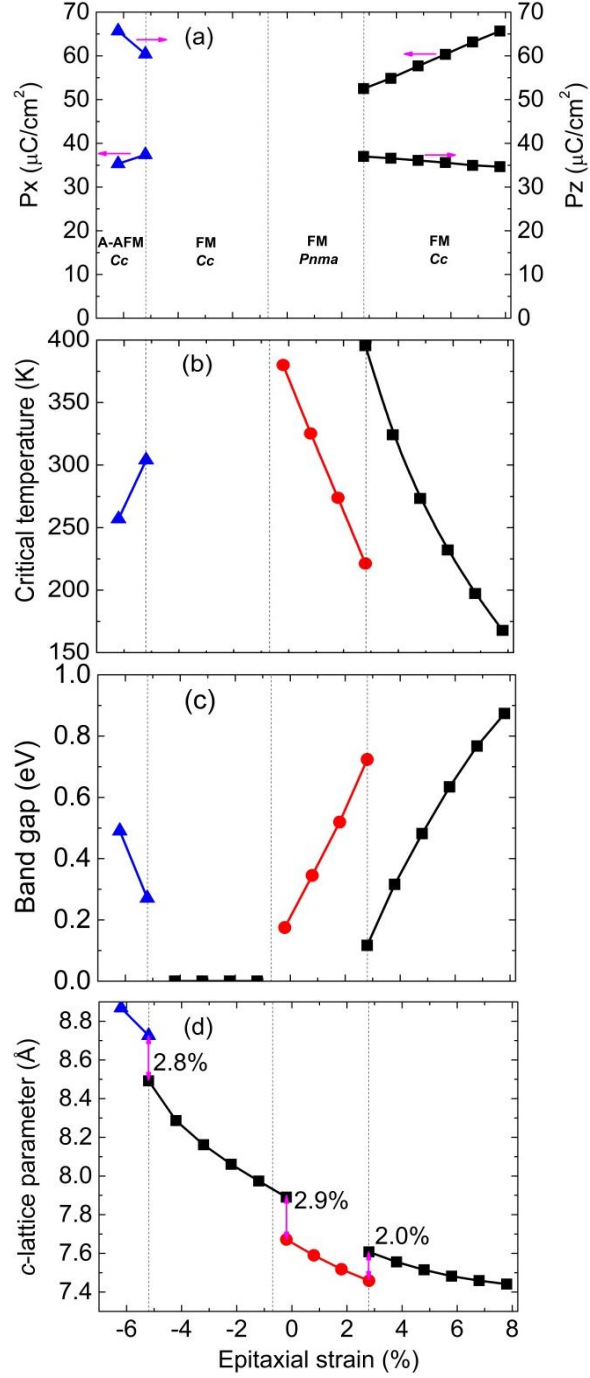

FIG. 2. Physical properties of epitaxial (001) BiMnO$_3$ thin film in the lowest energy state at each strain. (a) Calculated ferroelectric polarization (P$_y$ = P$_x$). (b) Magnetic critical temperatures of FM and A-AFM orders. (c) Band gap. (d) *c*-lattice parameter. Symbols are the same as those in Figure 1. The ground states (see Fig. 1) for each strain region are indicated by A-AFM *Cc*, FM *Cc*, FM *Pnma* and FM *Cc*,

respectively.

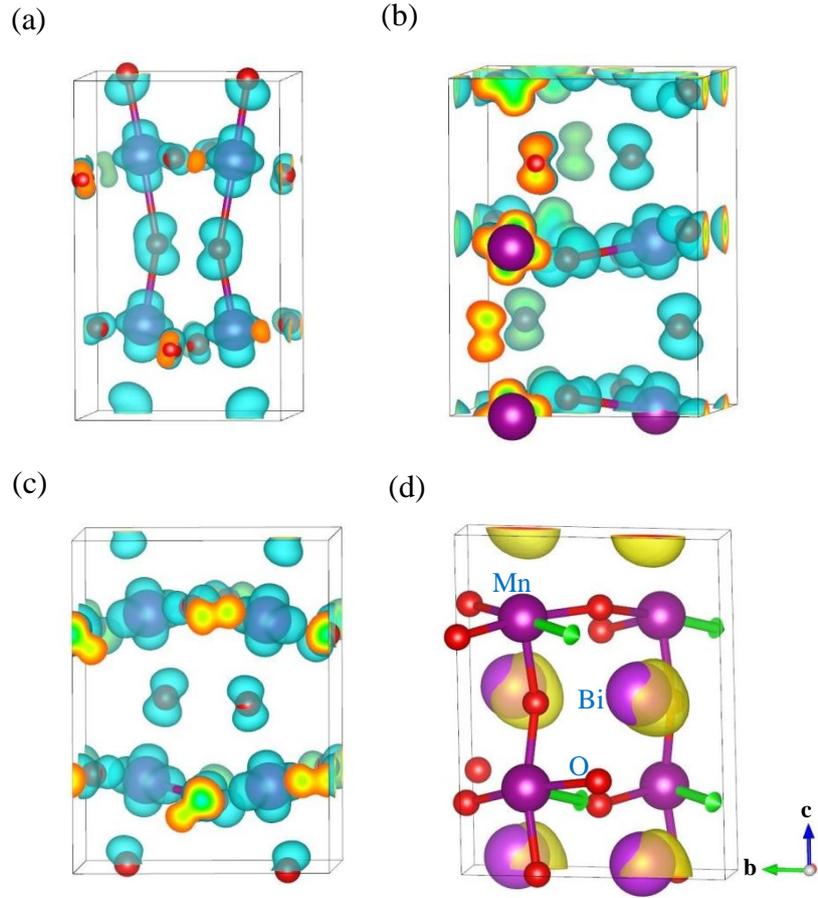

FIG. 3. Orbital orders in (a) A-AFM *Cc* structure at -5.2% strain, (b) FM *Pnma* structure at 0.8 % strain and (c) FM *Cc* structure at 3.8% strain. Only long Mn-O bonds are shown and Bi ion has been removed for clarity. (d) Isosurface of the electron localization function with the value of 0.8 for the FM *Cc* structure at 3.8% strain. The green arrow indicates the easy-axis of the $Mn^{3+}$ ion.

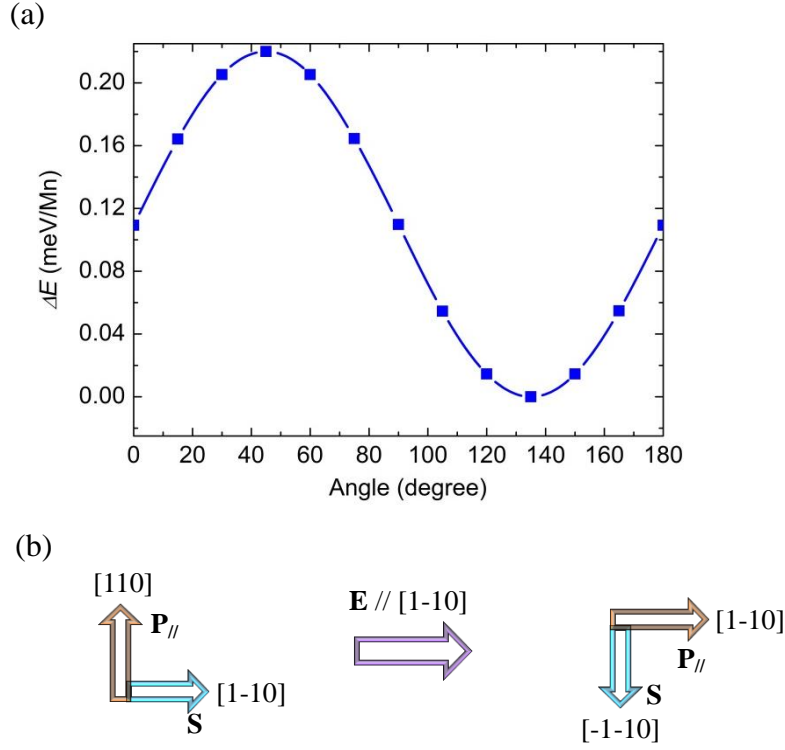

FIG. 4. (a) Total energy (Square dots) of the FM *Cc* structure at 3.8% strain from the DFT+U+SOC calculations as a function of the angle between the magnetization direction and the pseudocubic [100]-direction when the magnetization is rotated in the *ab*-plane. The line is the fitted result using the single-ion anisotropy model defined in the text. (b) Schematic illustration of the magnetoelectric coupling in the FM-FE *Cc* phase. **P**, **S** and **E** represent the FE polarization, Mn spin and external electric field, respectively. Here the directions are defined within the pseudocubic setting.

# Supplementary Materials for

# Epitaxially Strained BiMnO$_3$ Films: High-Temperature Robust Multiferroic Materials with Novel Magnetoelectric Coupling

X. Z. Lu[a)], X. G. Gong, and H. J. Xiang[b)]

Key Laboratory of Computational Physical Sciences (Ministry of Education), State Key Laboratory of Surface Physics, and Department of Physics, Fudan University, Shanghai 200433, P. R. China

a) 11210190025@fudan.edu.cn; b) hxiang@fudan.edu.cn

## 1. Dependence of results on the Hubbard U value

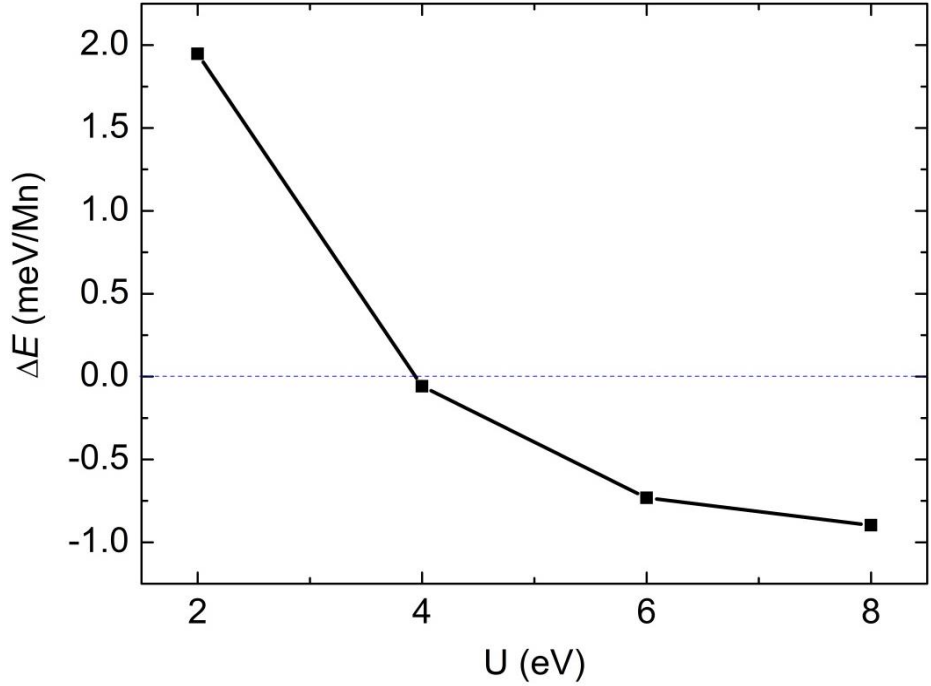

Figure S1. Energy difference between *C*2/*c* and *C*2 structures with the FM spin order versus the on-site repulsion U. When U is larger than 4 eV, the *C*2/*c* structure becomes more stable, in agreement with the experimental result[1]. In the main text, we mainly discussed the results obtained by using U = 4 eV, but our test calculations show that other appropriate U values do not change the main results (see below).

To see the effect of U on the results, we also perform the PBEsol+U calculations with U = 6 eV. We find that that the results with U = 6 eV show qualitatively agreement with those (with U = 4 eV) discussed in the main text. Quantitatively, the transition from the PE-FM *Pnma* phase to the FE-FM *Cc* phase shifts to somewhat higher strain. For the calculated polarization of *Cc* structure at 3.8% strain using U = 6 eV, it is also very close to that obtained by using U = 4 eV. The calculated FM critical temperatures at 3.8% strain are increased to be 417 K and 195 K within mean field theory and Monte Carlo simulation, respectively. Thus, our results in the main discussions should be valid.

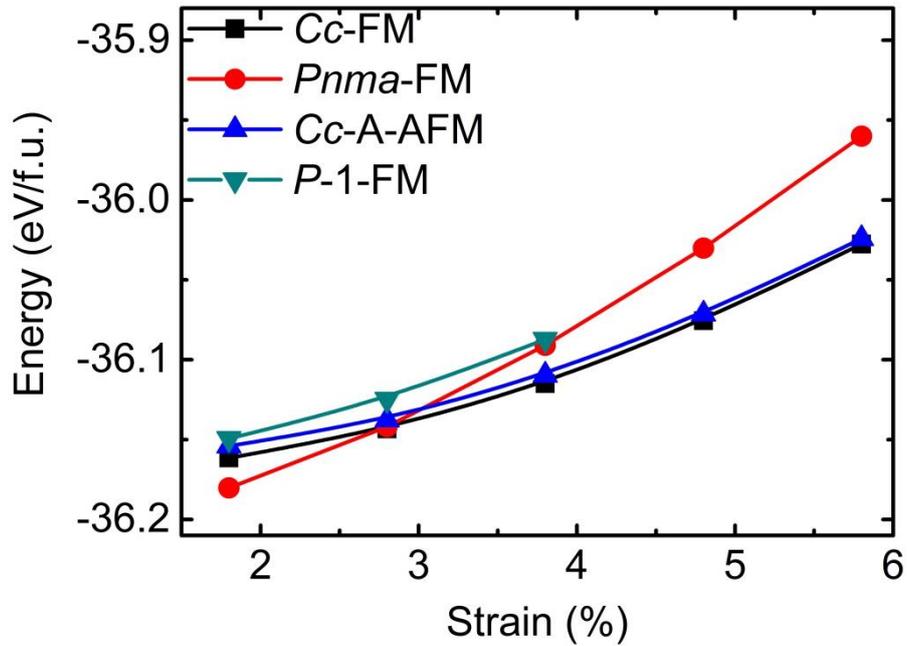

Figure S2. Calculated total energies versus the epitaxial strains (coherent epitaxy) for the various spin orders with *Pnma* and *Cc* space groups in the range of 1.8-5.8% using the same settings (such as LDA+U with U = 6 eV and J = 0.8 eV) for the DFT calculations as those adopted in ref 2.

As shown in Figure S2, the phase diagram in the range of 2.8-5.8% strain (the range was not studied in ref 2) is good agreement with the result obtained by using the settings used in the main text, which again indicates that our discussions in the main text are valid.

**2. Monte Carlo simulations for the magnetic critical temperatures using U = 4 eV**

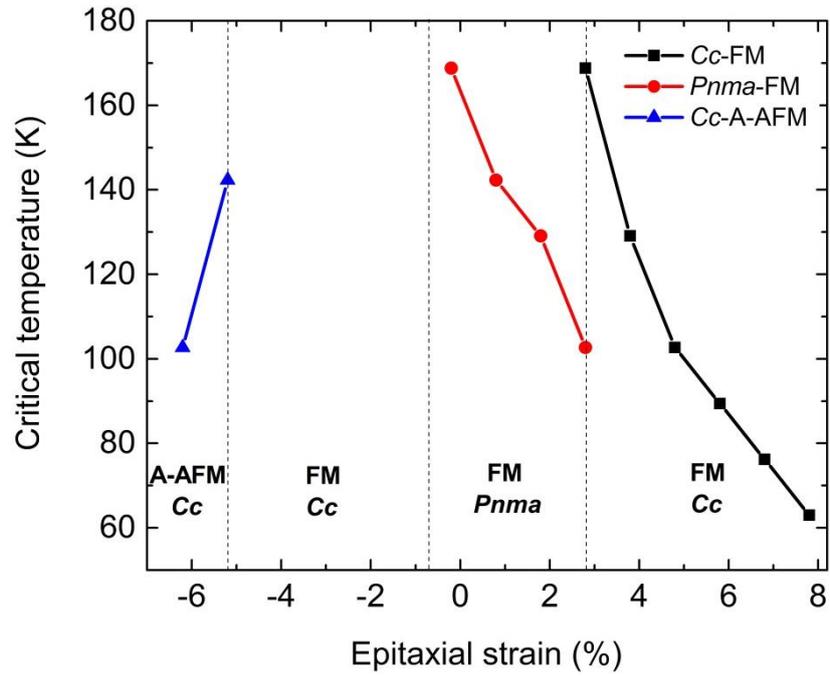

Figure S3. Magnetic critical temperatures of FM and A-AFM orders in the lowest energy structure at each strain obtained from our Monte Carlo simulations.

## 3. Values for the spin exchange parameters

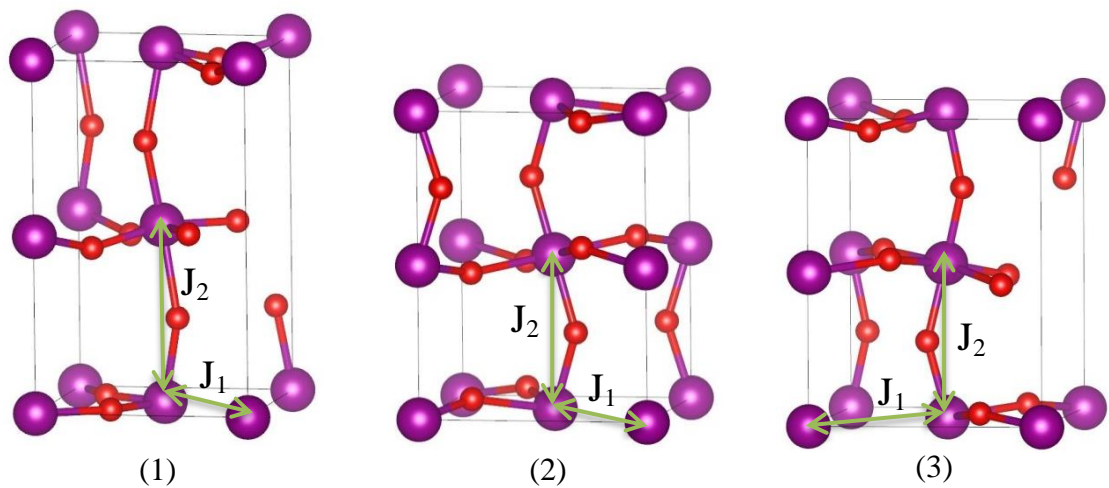

Figure S4. Representative structures. (1) A-AFM-FE *Cc* structure at -5.2% strain, (2) FM-PE *Pnma* structure at 0.8 % strain and (3) FM-FE *Cc* structure at 3.8% strain. The spin exchange paths $J_1$-$J_2$ in the selected structures are also shown. Bi ion has been removed for clarity.

Table SI. Calculated exchange parameters (J) deduced from GGA+U calculations, where $J_i$ is an effective spin exchange value obtained by setting $|S_i| = 1$, namely, $J_{ij}^{eff} = J_{ij}S_iS_j$ for a spin dimer ij.

| Strain (%) | $J_1$ (meV) | $J_2$ (meV) |
| --- | --- | --- |
| -6.2 | -3.293 | 15.550 |
| -5.2 | -7.122 | 11.949 |
| -0.2 | -13.086 | -6.583 |
| 0.8 | -11.335 | -5.359 |
| 1.8 | -9.628 | -4.339 |
| 2.8$^*$ | -7.634 | -3.799 |
| 2.8$^{**}$ | -15.458 | -3.158 |
| 3.8 | -12.964 | -2.006 |
| 4.8 | -11.138 | -1.277 |
| 5.8 | -9.596 | -0.808 |
| 6.8 | -8.290 | -0.417 |
| 7.8 | -7.170 | -0.113 |

Note: The superscript * represents the value for the FM *Pnma* structure and ** for the FM *Cc* structure.

## 4. Origin of the magnetic ground state

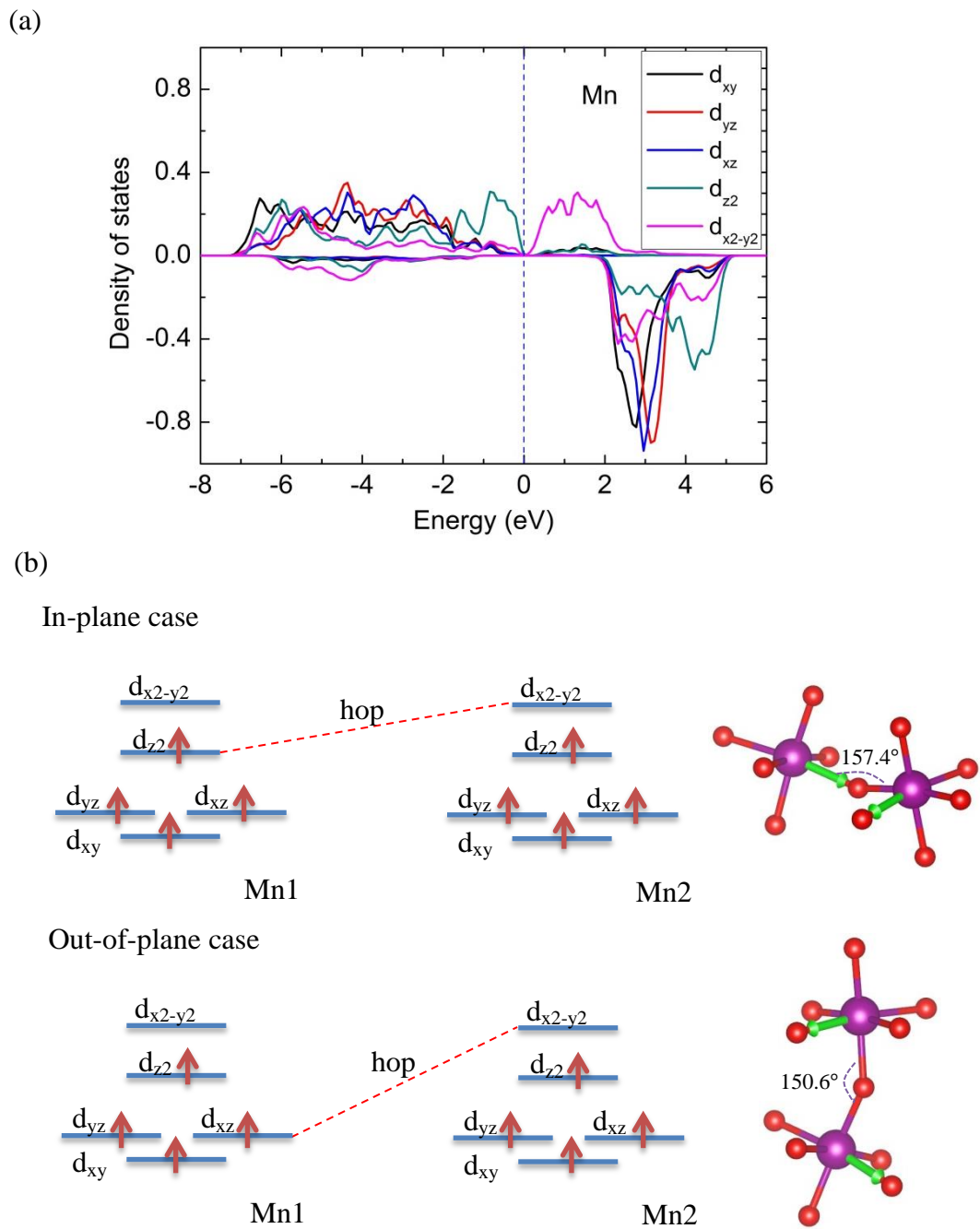

Figure S5. (a) Partial density of states for the *Cc* structure at 3.8% strain with a 0.1 eV broadening. The d orbitals refer to the local coordination system, where the z axis is along the long Mn-O bond direction. (b) Left panel: Schematic illustration of $\sigma - \sigma$ coupling and $\sigma - \pi$ coupling for the in-plane and out-of-plane cases, respectively. Right panel: Corresponding Mn1-O-Mn2 angles in the two cases. Green arrow

indicates the local z axis.

The results depicted in Figure S5 show that the coupling between the nearest neighbors of Mn ions in the in-plane is a strong σ − σ coupling between the occupied $d_{z2}$ orbital in Mn1 and the unoccupied $d_{x2-y2}$ orbital in Mn2, while a σ − π coupling takes place between the occupied $d_{yz}/d_{xz}$ orbitals in Mn1 and unoccupied $d_{x2-y2}$ orbital in Mn2 for the out-of-plane nearest-neighbors because of the small energy difference between them. Thus the magnetic order for the FM-FE *Cc* phase is ferromagnetic. The similar explanations are also expected for the A-AFM-FE *Cc* and FM-PE *Pnma* phases.

## 5. The calculated Born effective charges (in e)

Bi1:

| | | |
|---|---|---|
| 4.66586 | 0.08695 | 0.20733 |
| -0.23560 | 4.64745 | -0.75794 |
| -0.38205 | 0.03076 | 4.66746 |

Bi2:

| | | |
|---|---|---|
| 4.63177 | -0.24872 | -0.79407 |
| 0.06956 | 4.65297 | 0.17143 |
| 0.03076 | -0.38205 | 4.66746 |

Mn1:

| | | |
|---|---|---|
| 2.61889 | 0.24418 | -0.28726 |
| 0.41129 | 3.72095 | -0.69644 |
| -0.25105 | -0.04020 | 3.75128 |

Mn2:

| | | |
|---|---|---|
| 3.70967 | 0.40014 | -0.72010 |
| 0.23351 | 2.60728 | -0.31200 |
| -0.04020 | -0.25105 | 3.75128 |

O1:

| | | |
|---|---|---|
| -1.99326 | -0.23248 | 0.02432 |
| -0.37750 | -2.94542 | 0.23531 |
| 0.14128 | 0.06710 | -3.11099 |

O2:

| | | |
|---|---|---|
| -3.19214 | 0.06573 | -0.13837 |
| 0.02120 | -3.14581 | 1.28552 |
| -0.04866 | 0.39542 | -1.80354 |

O3:

| | | |
|---|---|---|
| -2.77319 | -0.02351 | 0.31879 |
| 0.07001 | -1.57778 | -0.13267 |
| -0.07075 | 0.15883 | -3.50357 |

O4:

|  |  |  |
|---|---|---|
| -1.61458 | 0.08258 | -0.12223 |
| -0.06035 | -2.76073 | 0.32953 |
| 0.15883 | -0.07075 | -3.50357 |

O5:

|  |  |  |
|---|---|---|
| -2.95491 | -0.40996 | 0.23765 |
| -0.24345 | -2.02477 | 0.02684 |
| 0.06710 | 0.14128 | -3.11099 |

O6:

|  |  |  |
|---|---|---|
| -3.10667 | 0.03072 | 1.27350 |
| 0.10649 | -3.18280 | -0.14916 |
| 0.39542 | -0.04866 | -1.80354 |

## 6. Phonon spectrum for the *Cc* structure at 3.8% strain

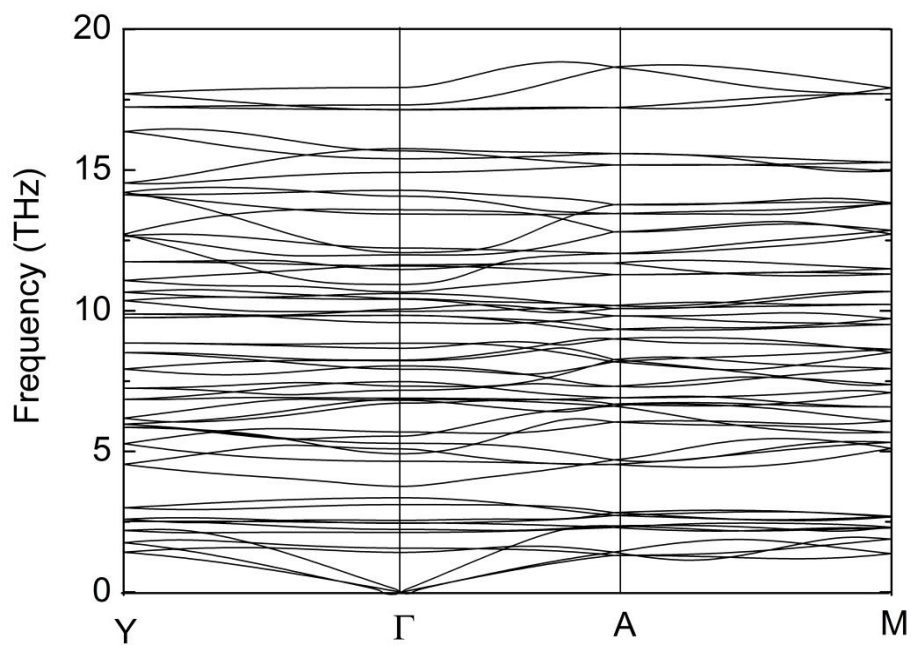

Figure S6. Phonon spectrum for the *Cc* structure at 3.8% strain. The phonon frequencies are calculated using the frozen phonon method in $2\sqrt{2} \times 2\sqrt{2} \times 4$ supercell with a $2 \times 2 \times 1$ *k*-point mesh at the Γ, A, M and Y points of Brillouin zone of 20-atom cell. The figure indicates that the structure is dynamically stable.

## 7. The total single-ion anisotropic energy of the FM *Cc* state at 3.8% strain

We compute the total single-ion anisotropic energy of the FM *Cc* state by using the following equation:

$$E_{SIA} = \sum_i A(\mathbf{S}_i \cdot \mathbf{n}_i)^2$$

$$= 2A(\sin^2\theta\,(n_{1x}^2 + n_{1y}^2) + 2\cos^2\theta\,n_{1z}^2 + 4\sin^2\theta\sin\varphi\cos\varphi\,n_{1x}n_{1y}$$

$$+ 2\sin\theta\cos\theta\,(\sin\varphi + \cos\varphi)(n_{1x} + n_{1y})n_{1z}) \tag{1}$$

where $E_{SIA}$ is the total single-ion anisotropic energy, $A = -3.105$ meV/Mn is the effective single-ion anisotropic parameter, $\mathbf{S}_i = \mathbf{S}(\theta, \varphi)$ represents the i-th $Mn^{3+}$ spin ($|\mathbf{S}_i| = 1$) where $\theta$ and $\varphi$ of the sphere coordinates determine the spin direction, and $\mathbf{n}_i$ the direction of its easy-axis where $n_{1x} = 0.036$, $n_{1y} = -0.984$ and $n_{1z} = -0.176$ are the Cartesian components of $\mathbf{n}_1$. To find the extreme value of $E_{SIA}$, we need to solve the following equations:

$$\begin{cases} \tan 2\theta = -\dfrac{2n_{1z}(n_{1x} + n_{1y})(\sin\varphi + \cos\varphi)}{n_{1x}^2 + n_{1y}^2 - 2n_{1z}^2 + 4\sin\varphi\cos\varphi\,n_{1x}n_{1y}} & (2) \\ \cos^2\varphi - \sin^2\varphi = -\dfrac{\cos\theta\,n_{1z}(n_{1x} + n_{1y})(\cos\varphi - \sin\varphi)}{2\sin\theta\,n_{1x}n_{1y}} & (3) \end{cases}$$

Situation (i): $\cos\varphi = \sin\varphi$, i.e., $\varphi = 45°$, then $\theta \approx 165°$ can be obtained by Eq. (2). In this case, $E_{SIA} = 0$ is the maximum value.

Situation (ii): $\cos\varphi \neq \sin\varphi$, then Eq. (3) becomes

$$\sin\varphi + \cos\varphi = -\frac{\cos\theta\,n_{1z}(n_{1x} + n_{1y})}{2\sin\theta\,n_{1x}n_{1y}} \tag{3'}$$

Since Eq. (2) and Eq. (3) are self-consistent, then the solutions are $\theta = 90°$ and $\varphi = 135°$. The result shows that $E_{SIA} \approx 2A$ is the minimum value. Therefore, the FM *Cc* state has a strong anisotropy with an energy difference of 1.55 meV/Mn between the hard magnetization case and the easy magnetization case.